\documentclass[%
reprint,
superscriptaddress,
 amsmath,amssymb,
 aps,
prx,
floatfix,
]{revtex4-2}

\usepackage{graphicx}
\usepackage{dcolumn}
\usepackage{bm}
\usepackage[hidelinks]{hyperref}

\usepackage[dvipsnames]{xcolor}
\hypersetup{
	colorlinks   = true, 
	urlcolor     = blue, 
	linkcolor    = blue, 
	citecolor   = blue 
}
\usepackage[T1]{fontenc} 
\usepackage{siunitx} 
\usepackage{soul} 
\usepackage{multirow} 
\usepackage[flushleft]{threeparttable}
\usepackage{mathptmx} 
\usepackage{cancel}

\providecommand{\abs}[1]{\lvert#1\rvert}

\graphicspath{ 
	{figures} 
}
\makeatletter
\def\input@path{
	{figures/} 
} 
\makeatother

\begin{document}

\preprint{APS/123-QED}

\title{Directional Emission of a Readout Resonator for Qubit Measurement}

\newcommand{\RLE}{\affiliation{Research Laboratory of Electronics, Massachusetts Institute of Technology, \\ Cambridge, Massachusetts 02139, USA}}
\newcommand{\EECS}{\affiliation{Department of Electrical Engineering and Computer Science, Massachusetts Institute of Technology, \\ Cambridge, Massachusetts 02139, USA}}
\newcommand{\LL}{\affiliation{MIT Lincoln Laboratory, Lexington, Massachusetts 02421, USA}}
\newcommand{\Harvard}{\affiliation{Harvard John A. Paulson School of Engineering and Applied Sciences, Harvard University, \\ Cambridge, Massachusetts 02138, USA}}
\author{Alec~Yen}\email{alecyen@mit.edu}\EECS\RLE
\author{Yufeng~Ye}\EECS\RLE
\author{Kaidong~Peng}\EECS\RLE
\author{Jennifer~Wang}\EECS\RLE
\author{Gregory~Cunningham}\RLE\Harvard
\author{Michael~Gingras}\LL
\author{Bethany~M.~Niedzielski}\LL
\author{Hannah~Stickler}\LL
\author{Kyle~Serniak}\RLE\LL
\author{Mollie~E.~Schwartz}\LL
\author{Kevin~P.~O'Brien}\email{kpobrien@mit.edu}\EECS\RLE

\date{\today}

\begin{abstract}

We propose and demonstrate transmission-based dispersive readout of a superconducting qubit using an all-pass resonator, which preferentially emits readout photons toward the output.
This is in contrast to typical readout schemes, which intentionally mismatch the feedline at one end so that the readout signal preferentially decays toward the output.
We show that this intentional mismatch creates scaling challenges, including larger spread of effective resonator linewidths due to non-ideal impedance environments and added infrastructure for impedance matching.
A future architecture using multiplexed all-pass readout resonators would avoid the need for intentional mismatch and potentially improve the scaling prospects of quantum computers.
As a proof-of-concept demonstration of ``all-pass readout,'' we design and fabricate an all-pass readout resonator that demonstrates insertion loss below 1.17 dB at the readout frequency and a maximum insertion loss of 1.53 dB across its full bandwidth for the lowest three states of a transmon qubit.
We demonstrate qubit readout with an average single-shot fidelity of 98.1\% in 600 ns; to assess the effect of larger dispersive shift, we implement a shelving protocol and achieve a fidelity of 99.0\% in 300 ns.

\end{abstract}

\maketitle

\newcommand{\cev}[1]{\reflectbox{\ensuremath{\vec{\reflectbox{\ensuremath{#1}}}}}}
\newcommand{\green}[1]{\color{ForestGreen} #1}
\newcommand{\red}[1]{\color{Red} #1}
\hyphenation{infeasible increasing unsustainable infrastructure mismatch input inconvenience JTWPAs JTWPA without unbounded}

\newcolumntype{P}[1]{>{\centering\arraybackslash}p{#1}}
\newcolumntype{M}[1]{>{\centering\arraybackslash}m{#1}}

\newcommand{\ci}{b_i}
\newcommand{\co}{b_o}
\newcommand{\li}{l_i}
\newcommand{\lo}{l_o}
\newcommand{\ri}{r_i}
\newcommand{\ro}{r_o}
\newcommand{\di}{c_i}
\newcommand{\doo}{c_o}
\newcommand{\ai}{a_i}
\newcommand{\aoo}{a_o}
\newcommand{\amode}{a}
\newcommand{\wa}{\wrr}
\newcommand{\ka}{\kr}
\newcommand{\Gl}{\Gamma_1}
\newcommand{\Gr}{\Gamma_2}
\newcommand{\wat}{\wrt}
\newcommand{\kat}{\krt}
\newcommand{\etaout}{\eta_{\text{out}}}

\newcommand{\Real}[1]{\mathrm{Re}\left\{#1\right\}}
\newcommand{\Imag}[1]{\mathrm{Im}\left\{#1\right\}}

\newcommand{\Gin}{\Gamma_\mathrm{in}}
\newcommand{\Gout}{\Gamma_\mathrm{out}}
\newcommand{\Gn}{\abs{\Gout}}

\newcommand{\VSWR}{\mathrm{VSWR}}
\newcommand{\kmax}{\kappa_\mathrm{max}}
\newcommand{\kmin}{\kappa_\mathrm{min}}

\newcommand{\krn}{\kappa_{r0}}
\newcommand{\krt}{\tilde{\kappa}_r}
\newcommand{\lr}{\lambda_r}
\newcommand{\eres}{\epsilon_\mathrm{eff,res}}
\newcommand{\efl}{\epsilon_\mathrm{eff,feed}}
\newcommand{\eeff}{\epsilon_\mathrm{eff}}
\newcommand{\vpfl}{v_{p,\mathrm{feed}}}
\newcommand{\vpr}{v_{p,\mathrm{res}}}

\newcommand{\aheo}{\left\{\begin{array}{l}
		\ahe \\
		\aho
	\end{array}\right\}}
\newcommand{\ahab}{\left\{\begin{array}{l}
		\aha \\
		\ahb
	\end{array}\right\}}
\newcommand{\ahba}{\left\{\begin{array}{l}
		\ahb \\
		\aha
	\end{array}\right\}}
\newcommand{\ainbin}{\left\{\begin{array}{l}
		\ain \\
		\bin
	\end{array}\right\}}
\newcommand{\binain}{\left\{\begin{array}{l}
		\bin \\
		\ain
	\end{array}\right\}}
\newcommand{\wreo}{\left\{\begin{array}{l}
		\wre \\
		\wro
	\end{array}\right\}}
\newcommand{\keo}{\left\{\begin{array}{l}
		\ke \\
		\ko
	\end{array}\right\}}
\newcommand{\aha}{{a}_1}
\newcommand{\ahda}{{a}_1^\dagger}
\newcommand{\ahb}{{a}_2}
\newcommand{\ahdb}{{a}_2^\dagger}
\newcommand{\bh}{{a}_q}
\newcommand{\bhd}{{a}_q^\dagger}
\newcommand{\sz}{{\sigma}_z}

\newcommand{\ke}{\kappa_e}
\newcommand{\ko}{\kappa_o}
\newcommand{\kr}{\kappa_r}
\newcommand{\ain}{{a}_\mathrm{in}}
\newcommand{\bin}{{b}_\mathrm{in}}
\newcommand{\aout}{{a}_\mathrm{out}}
\newcommand{\bout}{{b}_\mathrm{out}}

\newcommand{\gfix}{g_\text{total}}

\newcommand{\aho}{{a}_o}
\newcommand{\ahdo}{{a}_o^\dagger}
\newcommand{\ahe}{{a}_e}
\newcommand{\ahde}{{a}_e^\dagger}

\newcommand{\wrr}{\omega_r}
\newcommand{\wrra}{\omega_{r1}}
\newcommand{\wrrb}{\omega_{r2}}
\newcommand{\wq}{\omega_{01}}
\newcommand{\wret}{\tilde{\omega}_{e}}
\newcommand{\wre}{{\omega}_{e}}
\newcommand{\wrot}{\tilde{\omega}_{o}}
\newcommand{\wro}{{\omega}_{o}}
\newcommand{\wqt}{\tilde{\omega}_{01}}
\newcommand{\wrt}{\tilde{\omega}_{r}}

\newcommand{\Gp}{\Gamma_{P}}
\newcommand{\Tp}{T_{1,{P}}}
\newcommand{\chieff}{\chi_\text{eff}}

\newcommand{\ground}{|0\rangle}
\newcommand{\excited}{|1\rangle}

\newcommand{\wRO}{\omega_\mathrm{RO}}

\section{Introduction}

\label{sec:intro}
Dispersive readout \cite{blais_cavity_2004,schuster_ac_2005,gambetta_qubit-photon_2006}, broadband parametric amplifiers \cite{macklin_near-quantum-limited_2015,obrien_resonant_2014}, and Purcell filtering \cite{reed_fast_2010,sete_quantum_2015,jeffrey_fast_2014} have made it possible to perform multiplexed single-shot readout of superconducting qubit systems with high fidelity \cite{heinsoo_rapid_2018,krinner_realizing_2022}. 
Such capabilities will be critical for any application requiring real-time feedback, such as quantum error correction (QEC) \cite{acharya_suppressing_2023,chen_exponential_2021,krinner_realizing_2022}. 
In multiplexed readout, maximizing the collection of photons encoded with qubit information is critical to improving the signal-to-noise ratio (SNR). As a result, the feedline is often intentionally interrupted at one end (the ``input'') with a wideband mismatch with near-total reflection so that the readout signal preferentially decays from the resonators toward the detection apparatus (the ``output'') \cite{jeffrey_fast_2014,heinsoo_rapid_2018,krinner_realizing_2022,sank_system_2024}. In a transmission-based readout configuration, this intentional mismatch takes the form of an input capacitance \cite{heinsoo_rapid_2018,krinner_realizing_2022,jeffrey_fast_2014}; in a reflection-based readout configuration, one end of the feedline is terminated in an open \cite{sank_system_2024}.
However, we show that intentional mismatch creates standing waves that both significantly worsen the spread of resonator linewidths and require added infrastructure for impedance matching.
We detail these challenges in Sec.~\ref{sec:motivation}.
Despite its disadvantages, the intentional use of mismatch in the feedline has thus far been ubiquitous and essential to enable directional decay of the readout signal and maximize SNR \cite{heinsoo_rapid_2018,krinner_realizing_2022,sank_system_2024,jeffrey_fast_2014}.
Thus, an attractive yet unexplored research direction would investigate readout schemes that avoid intentional mismatch while preserving the directional decay of the readout signal.

Here, we present a proof-of-concept demonstration of ``all-pass readout,'' a transmission-based scheme in which the readout resonator demonstrates preferential directional emission of photons encoded with qubit information, thus avoiding the need for intentional mismatch.
In Sec.~\ref{sec:model}, we present a model for an all-pass resonator coupled to a transmon qubit. We experimentally demonstrate all-pass performance with the fabricated device detailed in Sec.~\ref{sec:device} and demonstrate high-fidelity single-shot readout in Sec.~\ref{sec:readout}.

\section{Limitations of Standing Waves in Readout}
\label{sec:motivation}

\begin{figure}
	\includegraphics{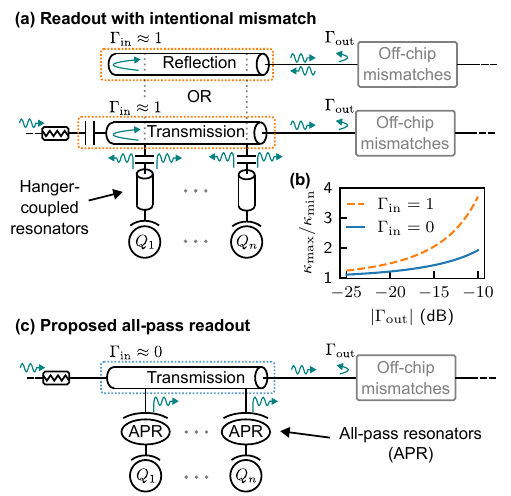}
	\caption{Spread in resonator linewidth $\kr$ due to off-chip mismatches. (a)~Multiplexed readout in reflection or in transmission using intentional mismatch to ensure preferential decay of the readout signal from resonators toward the output.
		(b)~The ratio of largest to smallest $\kr$ as a function of the output-side reflection coefficient.
		(c)~Proposed all-pass readout. No intentional mismatch is needed because the all-pass resonators preferentially emit readout photons in one direction.
	}
	\label{fig:kappa-span}
\end{figure}

Cyclic protocols for quantum error correction, such as the surface code, require reliable measurement of all data and syndrome qubits within a given cycle \cite{bengtsson_model-based_2024}.
As a result, robust control of the resonator linewidth $\kr$ is of critical importance. 
If $\kr$ is too large, the qubit lifetime can be shortened due to the Purcell effect \cite{reed_fast_2010,jeffrey_fast_2014,sete_quantum_2015}. 
Too small of $\kr$ is also problematic since the repetition rate of cyclic protocols is limited by the qubit with the slowest readout \cite{bengtsson_model-based_2024}.
In practice, it is common to see significant variation in the resonator linewidth; this has previously been attributed to off-chip impedance mismatches \cite{sank_system_2024,li_experimentally_2023,chow_characterizing_2015}. 
A multiplexed readout scheme with intentional mismatch is depicted in Fig.~\ref{fig:kappa-span}a, where hanger-coupled resonators are coupled to a feedline, and the intentional mismatch is represented by an input capacitance (in transmission \cite{heinsoo_rapid_2018,krinner_realizing_2022,jeffrey_fast_2014}) or as an open (in reflection \cite{sank_system_2024}.) 
In Fig.~\ref{fig:kappa-span}a, we depict resonators directly coupled to the feedline, but note that the variation of effective linewidth extends to designs with bandpass Purcell filters.
In the following analysis, we show that intentional mismatch worsens variation by increasing the sensitivity of the resonator linewidth to non-ideal impedance environments, such as off-chip mismatches and on-chip non-uniformities.

First, we show how intentional mismatch worsens the variation in resonator linewidth $\kr$ as a result of off-chip mismatch, where the spread is related to the voltage standing wave ratio, given by $\VSWR=\left(1+\abs{\Gout}\right)/\left(1-\abs{\Gout}\right)$. 
The output-side reflection coefficient $\Gout$ depends on the matching performance of off-chip components, such as the sample package, isolators, and filters.
For a system with intentional mismatch ($\Gin\approx1$), the ratio of the largest to smallest possible linewidths is given by the square of VSWR (see Appendix~\ref{sec:app-int-mismatch})
\begin{equation}
	\frac{\kmax}{\kmin}\left(\Gin=1\right)=\left(\frac{1+\abs{\Gout}}{1-\abs{\Gout}}\right)^2 = \VSWR^2.
	\label{eq:kspan1}
\end{equation}
This was confirmed experimentally in \cite{sank_system_2024}, where a spread in $\kr$ of $2\times$ was measured for $\abs{\Gout}=\SI{-16}{dB}$. 
From \eqref{eq:kspan1}, a clear way to mitigate spread in $\kr$ is to improve matching performance at the output.
As a complementary approach, we could also improve matching at the input by removing the intentional mismatch. 
If $\Gin=0$, the scaling of $\kmax/\kmin$ is now given by (see Appendix~\ref{sec:app-noint-mismatch}) 
\begin{equation}
	\frac{\kmax}{\kmin}\left(\Gin=0\right)=\frac{1+\abs{\Gout}}{1-\abs{\Gout}} = \VSWR.
	\label{eq:kspan2}
\end{equation}
As shown in Fig.~\ref{fig:kappa-span}b, this is a square-root reduction in $\kmax/\kmin$ compared to \eqref{eq:kspan1}.
The key advantage to this approach is that attenuators can be used for matching on the input side to minimize $\abs{\Gin}$ (see Fig.~\ref{fig:kappa-span}c).
In contrast, attenuators cannot typically be used on the output side, since minimization of pre-amplification losses is critical for maximizing measurement efficiency \cite{clerk_introduction_2010}.


Intentional mismatch also makes the resonator linewidth $\kr$ highly sensitive to non-uniformities in the effective permittivity $\eeff$ as seen by a coplanar waveguide across a chip.
Non-uniformities are especially prevalent in flip-chip processors, where inter-chip spacing variation and nonlinear chip deformations on the order of $\SI{1}{\micro m}$ can cause resonator frequencies to vary by a few percent across a chip \cite{li_experimentally_2023,kosen_building_2022}. 
Typical non-uniformities make multiplexing beyond 15 resonators challenging, even for a generous tolerance range of $\pm30\%$.
See Appendix~\ref{sec:app-nonuniform-full} for a detailed analysis.

As a final comment, we note that intentional mismatch also impedes the scaling of quantum computers by requiring the addition of nonreciprocal components to impedance match to downstream broadband amplifiers \cite{peng_x-parameter_2022, peng_floquet-mode_2022,heinsoo_rapid_2018}.
We refer the reader to Appendix~\ref{sec:app-amplifier-matching} for further details.

\section{Implementation of All-Pass Readout}
\label{sec:model}
To demonstrate the experimental feasibility of directional emission of a readout resonator, we develop a design for an all-pass resonator coupled to a transmon qubit.
In sharp contrast to an overcoupled hanger-coupled readout resonator, which exhibits total reflection on resonance, an all-pass readout resonator instead approaches total transmission on resonance ($|S_{21}|\rightarrow1$). We can understand how all-pass behavior arises by considering a resonator with both an even and an odd mode coupled to a waveguide (Fig.~\ref{fig:ckt-model}a). The transmission is given by (see  Appendix~\ref{sec:app-cmt})
\begin{equation}
	S_{21} = 1 - \frac{\frac{\kappa_e}{2}}{j\left(\omega-\wre\right)+\frac{\kappa_e}{2}}
	- \frac{\frac{\kappa_o}{2}}{j\left(\omega-\wro\right)+\frac{\kappa_o}{2}}\,,
	\label{eq:drr-S21}
\end{equation}
where $\wre$ ($\wro$) is the even (odd) mode frequency and  $\kappa_e$ ($\kappa_o$) is the even (odd) mode linewidth. The two modes will be degenerate if they have equal frequencies and equal linewidths, i.e., 
$\omega_e=\omega_o=\omega_{\text{deg}}$
and
$\kappa_{e} = \kappa_{o} = \kappa_{\text{deg}}$.
In this special case, \eqref{eq:drr-S21} becomes
\begin{equation}
	S_{21,\mathrm{deg}} = 1-\frac{\kappa_\text{deg}}{j\left(\omega-\omega_{\text{deg}}\right)+\frac{\kappa_\text{deg}}{2}}\,,
\end{equation}
which achieves the desired all-pass behavior \cite{manolatou_coupling_1999}.
This can be understood by noting that at $\omega=\omega_\mathrm{deg}$, we have $|S_{21,\mathrm{deg}}|=1$.
\begin{figure}
	\includegraphics{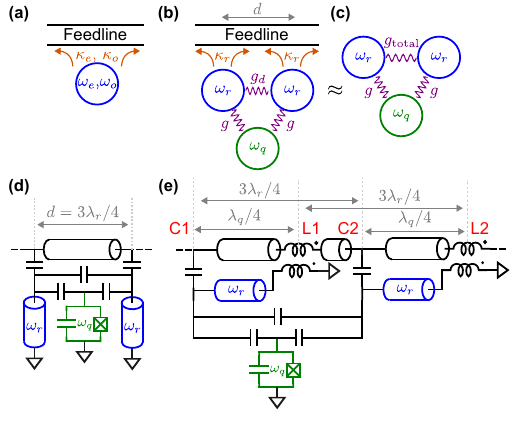}
	\caption{Models for all-pass readout.
		(a)~Even and odd resonant modes coupled to a feedline.
		(b)~Two identical resonators coupled to both a feedline and a qubit mode.
		(c)~Simplified representation of (b) where $\gfix$ is the sum of $g_{d}$ and the waveguide-mediated coupling $g_w$.
		(d)~A possible circuit implementation of (b), where $d=3\lambda_r/4$ for all-pass behavior.
		(e)~A modified version of (d) that includes Purcell suppression using interference of the qubit mode in the feedline \cite{yen_interferometric_2024}.}
	\label{fig:ckt-model}
\end{figure}

With these conditions for all-pass behavior in mind, we design a resonator with even and odd modes to perform readout of a qubit (see Fig.~\ref{fig:ckt-model}b). A transmon qubit mode $\bh$ with frequency $\wq$ and charging energy $E_C$ is coupled symmetrically  to two identical resonator modes $\aha$ and $\ahb$ with frequency $\wrr$. The resonators are coupled to a waveguide with linewidth $\kr$ and separated by distance $d$.  The qubit-resonator coupling rates have magnitude $g$, and the direct resonator-resonator coupling rate has magnitude $g_d$. We can sum the direct and waveguide-mediated coupling rates and simplify Fig.~\ref{fig:ckt-model}b as Fig.~\ref{fig:ckt-model}c, where $\gfix=g_d+g_w$ (see Appendix~\ref{sec:app-cmt}).
The Hamiltonian of Fig.~\ref{fig:ckt-model}c is approximately represented in the Fock basis by 
\begin{equation}
	\begin{aligned}
		{H}  &=  \wrr\ahda\aha + \wrr\ahdb\ahb + \wq\bhd\bh - \frac{E_C}{2}\bhd\bhd\bh\bh \\
		&+ \gfix\left(\ahda+\aha\right)\left(\ahdb+\ahb\right) \\
		&+ g\left(\ahda+\aha\right)\left(\bhd+\bh\right)
		+ g\left(\ahdb+\ahb\right)\left(\bhd+\bh\right).
	\end{aligned}
	\label{eq:symHfullmaintext}
\end{equation}
After a transformation of basis and truncation of the transmon to the first two energy levels (see Appendix~\ref{sec:app-hamiltonian}), we can derive the dispersive Hamiltonian
\begin{equation}
	\begin{aligned}
		{\tilde{H}}_{\mathrm{disp}}
		& = \frac{\wqt}{2}\sz + \wre \ahde\ahe + \wro\ahdo\aho\,,
	\end{aligned}
\end{equation}
where the dressed qubit frequency $\wqt$, even mode frequency $\wre$, odd mode frequency $\wro$, and effective total dispersive shift $2\chi_{01}$ are approximated by
\begin{equation}
	\wqt=\wq+\frac{2g^2}{\Delta}\,,
\end{equation}
\begin{equation}
	\wre=\wrr+\gfix-\frac{2g^2}{\Delta-E_C}+2\chi_{01}\sz\,,
\end{equation}
\begin{equation}
	\wro=\wrr-\gfix\,,
\end{equation}
and
\begin{equation}
	\chi_{01}=-\frac{g^2E_C}{\Delta\left(\Delta-E_C\right)}\,,
	\label{eq:chi}
\end{equation}
where we define the qubit-resonator detuning as $\Delta=\wq-\wrr$.
Note that the even mode experiences a total dispersive shift of $4\chi_{01}$, whereas the odd mode does not.
This is because the qubit is positioned at the node of the odd mode of this system.
Thus, the effective  total dispersive shift is $2\chi_{01}$ (see Appendix~\ref{sec:app-chikappa}).

Recall that all-pass behavior is obtained when the even and odd modes are degenerate. 
Neglecting the dispersive shift for the moment, we see that $\wre=\wro$ when $\gfix=g^2/\left(\Delta-E_C\right)$. 
Intuitively, all-pass behavior occurs when the qubit-mediated coupling cancels the other channels of coupling between the two resonators. 
This degeneracy condition is analogous to that in the directional emission of an itinerant photon, where two qubit modes are made degenerate \cite{gheeraert_programmable_2020,kannan_-demand_2023}.
In our case, there is an optimal qubit frequency for which all-pass behavior occurs.
However, we note that the nonlinear dispersive shift of the qubit will break the degeneracy condition $\wre=\wro$. 
As a result, we choose to design the qubit in the small $2\chi_{01}/\kr$ regime to preserve all-pass behavior for relevant states in the computational subspace. 
See Appendix~\ref{sec:app-chikappa} for more details. 
As we will demonstrate in Sec.~\ref{sec:readout}, the design choice of large $\kr$ enables us to still achieve high-fidelity readout in \SI{300}{ns}. 

A possible circuit implementation of Fig.~\ref{fig:ckt-model}b is depicted in Fig.~\ref{fig:ckt-model}d. Two $\lambda_r/4$ resonators are capacitively coupled to a transmon qubit and each other. These resonators are capacitively coupled to a feedline and separated by phase delay $\phi=2\pi d/\lambda_r=1.5\pi$ to match linewidths $\ke=\ko$ (see Appendix~\ref{sec:app-cmt}).
In practice, one benefits from including Purcell suppression to achieve fast readout \cite{reed_fast_2010,jeffrey_fast_2014,sete_quantum_2015}. 
To do so, we utilize Purcell suppression based on interference of the qubit mode in the feedline \cite{yen_interferometric_2024}, where each resonator is coupled to two spatially separate points on the feedline (Fig.~\ref{fig:ckt-model}e).
This is similar to the creation of subradiant states in giant atoms \cite{kockum_decoherence-free_2018,kannan_waveguide_2020}. 
For $\lambda_r/4$ coplanar waveguide resonators, we realize Purcell suppression by coupling the open end capacitively and the shorted end inductively to the feedline. The coupling points are spaced apart at $\lambda_q/4$ separation such that the qubit mode destructively interferes in the feedline. The resonators' respective capacitive (inductive) coupling points are then spaced apart by $3\lambda_r/4$, as needed for matching even- and odd-mode linewidths. 

Like traditional multiplexing schemes, multiplexed all-pass resonators would be spaced apart in frequency---several multiples of the resonator linewidth---to minimize measurement errors and frequency overlap between different readout units \cite{jeffrey_fast_2014,heinsoo_rapid_2018}.
For all-pass readout, this spacing serves the additional purpose of preserving the degeneracy condition and minimizing the interaction between resonators of different readout units.
This enables modular design, as each unit can be optimized separately.
See Appendix~\ref{sec:app-multiplex} for an analysis of the directionality of cascaded all-pass readout units as a function of frequency spacing.

\section{Device Description and Characterization}
\label{sec:device}
\begin{figure}[t]
	\includegraphics[width=\columnwidth]{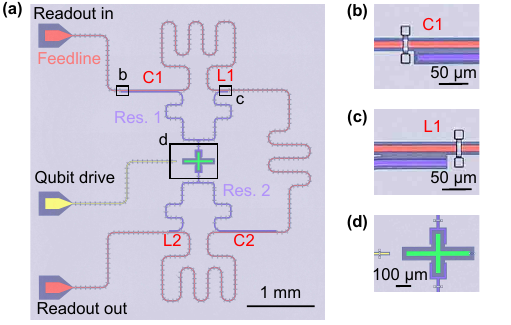}
	\caption{Experimental device. 
		(a)~False-colored chip micrograph with $\lambda_r/4$ transmission line resonators (blue), feedline (red), transmon qubit (green), and qubit drive line (yellow). 
		Zoomed images of (b)~the open end of the $\lambda_r/4$ resonator, (c)~the shorted end of the $\lambda_r/4$ resonator, and (d)~symmetric coupling of transmon qubit. Labels denoting where the resonator is capacitively (C1, C2) and inductively (L1, L2) coupled correspond to Fig.~\ref{fig:ckt-model}e.}
	\label{fig:micrograph}
\end{figure}

\begin{figure}[t]
	\includegraphics{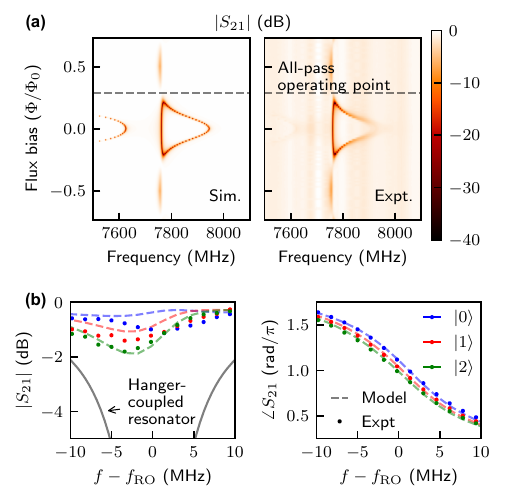}
	\caption{Transmission of device.
		(a)~Comparison of simulated and measured $|S_{21}|$ as a function of flux bias and frequency with the transmon in the ground state. Finite-element simulation approximates the transmon as a linear oscillator. 
		(b)~Measured magnitude and phase of $S_{21}$ versus frequency at the all-pass operating point for the three lowest states of the transmon. We overlay predicted curves using an analytic model. The $|S_{21}|$ of a conventional readout resonator (no intentional mismatch) with equivalent $\kr$ is shown for comparison.}
	\label{fig:drr_S21}
\end{figure}

We experimentally realize an all-pass readout resonator. The chip micrograph is shown in Fig.~\ref{fig:micrograph}a, corresponding to the circuit model in Fig.~\ref{fig:ckt-model}e, with the addition of a superconducting quantum interference device (SQUID) in place of a single Josephson junction.
The identical $\lambda_r/4$ resonators (blue) are coupled to the feedline (red) capacitively on the open end (Fig.~\ref{fig:micrograph}b) and inductively on the shorted end (Fig.~\ref{fig:micrograph}c). 
The claw design in Fig.~\ref{fig:micrograph}d provides symmetric qubit-resonator coupling rate $g$ to a single-island flux-tunable transmon qubit (green) with global magnetic flux bias and dedicated drive line (yellow).
We minimize sensitivity to spatial variation by making the two quarter-wave resonators geometrically identical and in very close proximity on the chip.

A 2D sweep of $|S_{21}|$ versus flux bias and frequency is shown in Fig.~\ref{fig:drr_S21}a. 
This tunable flux bias enables us to \textit{in situ} tune the qubit-mediated coupling between the resonators, similar to the qubit-mediated coupling used for tunable couplers \cite{yan_tunable_2018}. 
We observe that all-pass behavior occurs near the flux bias $|\Phi/\Phi_0|=0.291$, corresponding to a dressed qubit frequency of $\wqt/2\pi=\SI{6086}{MHz}$, and choose this as our operating point. 
In Fig.~\ref{fig:drr_S21}a, we show agreement with finite-element simulation, where the transmon qubit is approximated as a linear oscillator. 
See Appendix~\ref{sec:app-fem} for more details.
We note that near $\Phi/\Phi_0=0$, due to the strong hybridization of the resonator and qubit modes, the measured features are fainter than in simulation. This is because simulation approximates the qubit as a linear oscillator, whereas in reality, the qubit's nonlinearity prevents the population of higher excited states by microwave drives.

The measured $S_{21}$ magnitude and phase for the lowest three transmon states are shown in Fig.~\ref{fig:drr_S21}b. We calibrate the $S_{21}$ with respect to a through line that bypasses the package and wire-bonded chip. Based on the average $|S_{21}|$ away from the resonant point, we estimate insertion loss due to the package of \SI{0.28}{dB} in the frequency range of interest (see Appendix~\ref{sec:app-setup}). Adjusting for this package loss, the all-pass resonator's transmission $|S_{21}|$ at the readout tone $\omega_{RO}/2\pi=\SI{7760.2}{MHz}$ is above $\SI{-1.17}{dB}$ for the lowest three transmon states. The lowest dip in $|S_{21}|$ is $\SI{-1.53}{dB}$.
From the measured phase, we find a bare resonator frequency $\wrr/2\pi=\SI{7756.4}{MHz}$, resonator linewidth $\kr/2\pi=\SI{14.5}{MHz}$, and dispersive shift $2\chi_{01}/2\pi=\SI{-1.10}{MHz}$. From \eqref{eq:chi}, we calculate a qubit-resonator coupling rate of $g/2\pi=\SI{93.4}{MHz}$.

In Fig.~\ref{fig:drr_S21}b, we overlay the analytic model for the full magnitude and phase of $S_{21}$, where we include loss due to the package. The analytic model numerically solves the system Hamiltonian in \eqref{eq:symHfullmaintext} for the eigenmodes $\wre$ and $\wro$ and substitutes into the expression for $S_{21}$ in \eqref{eq:drr-S21}. Higher excited states of the transmon qubit and resonators are included to the point that the calculation converges. The linewidths of the system are given by $\ke=\kr\left(1+\cos\phi\right)$ and $\ko=\kr\left(1-\cos\phi\right)$ (see Appendix~\ref{sec:app-cmt}). We use the analytic model to estimate the two remaining unknowns of the system, which are the phase delay $\phi=1.55\pi$ and fixed resonator-resonator coupling rate $\gfix/2\pi=\SI{-5.1}{MHz}$. The even- and odd-mode linewidths are then $\ke/2\pi=\SI{17.1}{MHz}$ and $\ko/2\pi=\SI{11.9}{MHz}$, within 10\% of expected values from finite-element simulation (see Appendix~\ref{sec:app-fem}). In a future design, the linewidths could be matched by ensuring $\phi=1.5\pi$.

Using finite-element simulation, we estimate a Purcell-limited lifetime of $\SI{70}{\micro s}$ for our device.
Indeed, we measure a prolonged qubit lifetime of $T_{01}=\SI{16}{\micro s}$ that we predict to be limited by the drive line and intrinsic decay of the qubit. 
Compared to finite-element simulation of a qubit coupled to two resonators with separate loss channels, our device achieves a Purcell suppression factor of $\sim\!\!10$ at the all-pass operating point, where $\wq/2\pi=\SI{6086}{MHz}$.
We note that this suppression could be substantially improved by aligning the notch frequency of the interferometric Purcell filter with the target $\wq$; at the filter's notch frequency near $\SI{4900}{MHz}$, the simulated Purcell suppression exceeds three orders of magnitude, of similar magnitude to a separate demonstration of interference-based suppression in \cite{yen_interferometric_2024}.
In practice, however, due to the estimated $T_1$ limit from intrinsic decay and the drive line, alignment of the notch frequency with the target $\wq$ would only result in a factor of 2 increase in $T_1$.
See Appendix~\ref{sec:app-ipf} for further details.

\section{Single-Shot Readout}
\label{sec:readout}
\begin{figure}
	\includegraphics{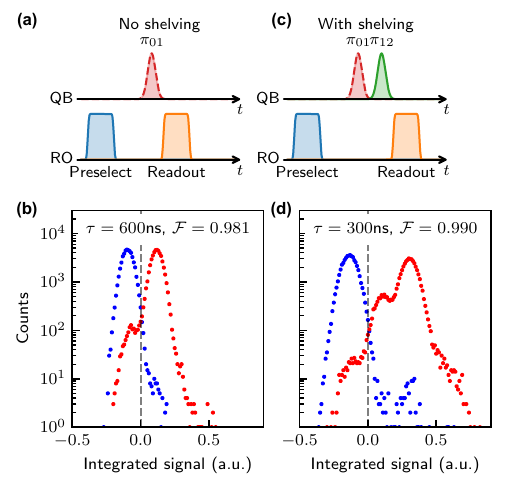}
	\caption{Single-shot readout. Pulse sequence (a,c) and histograms (b,d) of integrated signal when preparing qubit in the ground (blue) and excited (red) state, without and with shelving protocol. }
	\label{fig:readout}
\end{figure}

We demonstrate high-fidelity single-shot dispersive readout of the transmon qubit.
We perform a standard readout pulse sequence (Fig.~\ref{fig:readout}a).
The readout pulse is implemented as a flat-top pulse with Gaussian rising and falling edges of width \SI{50}{ns}.
We set the integration time $\tau$ to be the same as the duration that the readout pulse is at its maximum value.
The $\pi_{01}$ pulse is implemented as a DRAG pulse \cite{motzoi_simple_2009} with $2\sigma=\SI{20}{ns}$.  
A Josephson traveling-wave parametric amplifier \cite{macklin_near-quantum-limited_2015,obrien_resonant_2014} is used to amplify the output probe signal. 
The optimal drive strength, or photon number, is optimized from the tradeoff between minimizing measurement-induced state transitions and maximizing readout fidelity \cite{jeffrey_fast_2014,sank_measurement-induced_2016}.
A preselection pulse is included to only consider samples for which the qubit begins in the ground state. 
The preselection threshold is set at  99\% of the fitted cumulative Gaussian distribution of the ground state \cite{walter_rapid_2017}. 
There is a gap of  \SI{400}{ns} between the preselection pulse and the $\pi_{01}$ pulse.

Histograms of the amplified integrated quadratures are shown in Fig.~\ref{fig:readout}b, where the qubit has been prepared in either $|0\rangle$ or $|1\rangle$. 
We set the discrimination threshold at 0. 
We calculate the qubit readout assignment fidelity $\mathcal{F}=1-[P(0|1)+P(1|0)]/2$ \cite{sunada_fast_2022,chen_transmon_2023}, where $P(x|y)$ is the probability that the measurement outcome is $|x\rangle$ given that the qubit is prepared in the $|y\rangle$ state.
For integration time $\tau=\SI{600}{ns}$, we achieve readout fidelity of $\mathcal{F} = 98.1\%$. 
We obtain an error of preparing a qubit in the ground state of $P(1|0)=0.8\%$.
The error of preparing a qubit in the excited state is $P(0|1)=3.0\%$, where the error is dominated by qubit decay, near the theoretical estimate of infidelity due to qubit decay of $1-e^{-\tau/T_{01}}\approx3.7\%$.
Using measurement-induced dephasing \cite{gambetta_qubit-photon_2006}, we estimate that the optimized readout power corresponds to a steady-state resonator population of $\sim\!\!35$ photons.

To assess the effect of a larger effective dispersive shift and longer qubit lifetime, we repeat this experiment with a shelving protocol (Fig.~\ref{fig:readout}c), applying an unconditional $\pi_{12}$ pulse before the readout pulse \cite{chen_transmon_2023}.  
A qubit in the first excited state is transferred to the second excited state, whereas a qubit in the ground state is unaffected. 
The $\pi_{12}$ pulse is also implemented as a DRAG pulse; the pulse parameters are optimized similarly to the standard method for the $\pi_{01}$. 
The qubit is first prepared in $|1\rangle$, and Rabi experiments are performed to optimize pulse amplitude and frequency \cite{chen_transmon_2023}.

With this shelving protocol, the samples in the $|1\rangle$ state are excited to the $|2\rangle$ state, both increasing the dispersive shift from $2\chi_{01}/2\pi=\SI{-1.10}{MHz}$ to $2\chi_{02}/2\pi=\SI{-1.86}{MHz}$ and increasing the effective lifetime from $T_{01}=\SI{16}{\micro s}$ to $T_{02}=\SI{25}{\micro s}$.
The optimized readout power corresponds to a photon number of $\sim\!\!60$.
We achieve an improved readout fidelity of $\mathcal{F} = 99.0\%$ with integration time $\tau=\SI{300}{ns}$, with $P(1|0)=0.7\%$ and $P(0|1)=1.2\%$, in agreement with the estimate of infidelity due to qubit decay of $1-e^{-\tau/T_{02}}\approx1.2\%$.
As expected, the $P(0|1)$ error has decreased due to the shorter integration time and longer effective lifetime. 
In the histogram in Fig.~\ref{fig:readout}d, a ``shoulder'' representing the decay from the $|2\rangle$ to the $|1\rangle$ state is visible.

\section{Conclusions}

We have presented the first demonstration of qubit readout using a readout resonator which preferentially emits photons in one direction across its full bandwidth. 
We have shown that the scheme is experimentally feasible, achieving high-fidelity single-shot readout of a transmon qubit while preserving high directionality.

To enable faster readout, further investigation is needed of all-pass resonator structures that operate in the regime where the dispersive shift is comparable to the resonator linewidth, without sacrificing directionality in either of the qubit's computational states.
Another future direction would be to investigate structures  insensitive to local fabrication defects that could cause the bare resonator modes to be mismatched.
Future work may also perform analysis similar to \cite{probst_efficient_2015} to further characterize properties of the all-pass resonator.

Due to the larger footprint, we expect that all-pass readout will be best suited for 3D integration such as a flip-chip geometry, where the qubits are on a separate chip from the readout and control lines \cite{rosenberg_3d_2017}.
The design could be compacted by using closer spacing between resonators, tighter meandering, and increased effective dielectric constant, the last of which is possible through 3D integration \cite{rosenberg_3d_2017}.

Multiplexed qubit readout using readout resonators with directional emission would avoid the need for intentional mismatch, minimizing the spread of resonator linewidths and removing the overhead associated with impedance matching.
Moreover, the removal of intentional mismatch enables modular and flexible placement of resonator-qubit blocks along the feedline.
Future work could implement a multiplexed system using all-pass readout and empirically demonstrate lower variation of resonator linewidths.
Such research directions could potentially improve the scaling prospects of large-scale quantum computers.

\section*{Acknowledgements}
The authors thank Neereja M. Sundaresan, Oliver E. Dial, and David W. Abraham for insightful discussions.

This work was supported in part by the MIT-IBM Watson AI Lab. This material is based upon work supported by the Under Secretary of Defense for Research and Engineering under Air Force Contract No. FA8702-15-D-0001. Any opinions, findings, conclusions or recommendations expressed in this material are those of the author(s) and do not necessarily reflect the views of the Under Secretary of Defense for Research and Engineering. A.Y. and J.W. acknowledge support from the NSF Graduate Research Fellowship. Y.Y. acknowledges support from the IBM PhD Fellowship and the NSERC Postgraduate Scholarship. G.C. acknowledges support from the Harvard Graduate School of Arts and Sciences Prize Fellowship.

A.Y. developed the theoretical framework, designed the device and experimental procedure, conducted the measurements, and analyzed the data.
A.Y. and K.P.O. proposed the idea of all-pass readout.
A.Y., Y.Y., K.P., J.W., and G.C. contributed to the experimental setup.
M.G., B.M.N., and H.S. fabricated the device.
A.Y. and K.P.O. designed the custom package and circuit board.
A.Y., J.W., and K.P.O. developed the fabrication process for the custom circuit board.
A.Y. packaged the device.
K.S., M.E.S., and K.P.O. supervised the project.
A.Y. wrote the manuscript with input from all co-authors.
All authors contributed to the discussion of the results and the manuscript.

\appendix
\section{Sample and Setup}
\label{sec:app-setup}
\begin{figure}
	\includegraphics{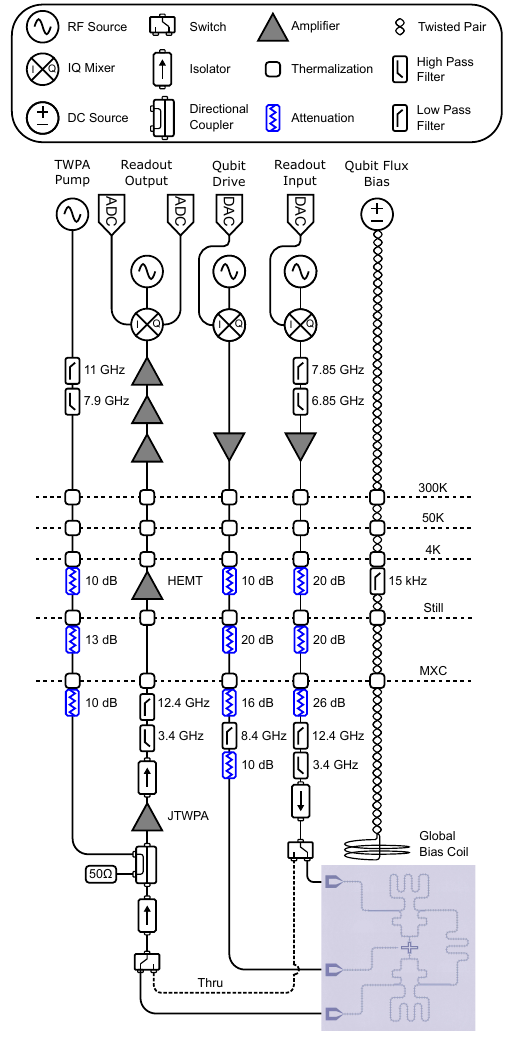}
	\caption{Experimental diagram and wiring.}
	\label{fig:fridge-diagram}
\end{figure}

\begin{table}
	\centering
	\caption{Measured device parameters.}
	\begin{tabular}{ l c  c}
		\hline
		\hline
		$|1\rangle\!\!-\!\!|0\rangle$ transition frequency & $\wqt/2\pi$ & \SI{6086}{MHz}\\
		$|2\rangle\!\!-\!\!|1\rangle$ transition frequency & $\tilde{\omega}_{12}/2\pi$ & \SI{5885}{MHz}\\
		$|1\rangle\!\!-\!\!|0\rangle$ relaxation time & $T_{01}$ & \SI{16}{\micro s}\\
		$|1\rangle\!\!-\!\!|0\rangle$ dephasing time & $T_{2,\mathrm{Echo}}$ & \SI{6}{\micro s}\\
		$|2\rangle\!\!-\!\!|0\rangle$ relaxation time & $T_{02}$ & \SI{25}{\micro s}\\
		Flux bias & $|\Phi/\Phi_0|$ & 0.291 \\ 
		Resonator frequency (bare) & $\wrr/2\pi$ & \SI{7756.4}{MHz} \\
		Resonator linewidth & $\kr/2\pi$ & \SI{14.5}{MHz}\\
		\multirow{2}*{Resonator dispersive shifts  \bigg\{ } & $2\chi_{01}/2\pi$ & \SI{-1.10}{MHz}\\ 
		& $2\chi_{02}/2\pi$ & \SI{-1.86}{MHz}\\
		\hline
		\hline
	\end{tabular}
	\label{tab:params}
\end{table}

\begin{figure}
	\includegraphics{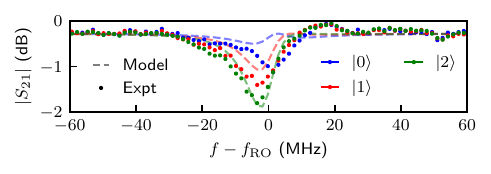}
	\caption{Measured $|S_{21}|$ away from the resonant point to estimate package loss. The analytic model includes the estimated loss.
	}
	\label{fig:app-plot-S21mag}
\end{figure}

\begin{table}
	\centering
	\caption{Measured transmission characteristics of all-pass readout resonator. Table values do not include an estimated package loss of 0.28~dB.}
	\begin{tabular}{ l c c  c c}
		\hline
		\hline
		& & $|0\rangle$ & $|1\rangle$ & $|2\rangle$ \\
		\hline
		\multirow{2}*{At readout tone $\omega_\mathrm{RO}$} & $|S_{21}|$    &\SI{-0.71}{dB} & \SI{-0.94}{dB} & \SI{-1.17}{dB}\\
		& $\angle S_{21}$    & $202^\circ$ & $188^\circ$& $179^\circ$\\
		Minimum of $|S_{21}|$  & & \SI{-0.85}{dB} & \SI{-1.13}{dB} & \SI{-1.53}{dB}\\
		\hline
		\hline
	\end{tabular}
	\label{tab:S21}
\end{table}

The transmon qubit and resonators are comprised of layers of thin-film aluminum on a silicon substrate. Airbridges are patterned to mitigate the formation of slot-line modes. The measured device parameters are listed in Table~\ref{tab:params}. We note that the reduced $T_2$ is expected, as we are far detuned from the flux-insensitive sweet spot at $\abs{\Phi/\Phi_0}=0$ \cite{koch_charge-insensitive_2007}.

The diagram of the experimental setup is shown in Fig.~\ref{fig:fridge-diagram}. We conducted the experiment in a Bluefors LD400 dilution refrigerator with a base temperature of \SI{15}{mK} at the mixing chamber (MXC). The device is enclosed by a superconducting aluminum shield, which is nested within a Cryoperm shield mounted at the MXC. 
To mitigate the presence of high-energy photons, low-pass filters are included at the qubit drive line (VLF-8400+) and readout input and output lines (RLC-F-30-12.4).
Microwave readout and drive tones are applied by a QICK ZCU111 RFSoC FPGA \cite{stefanazzi_qick_2022}.
The sample is housed within a custom package that is designed to suppress modes below \SI{17}{GHz}, which is more than sufficient for the experiment to avoid deleterious effects on qubit lifetime \cite{huang_microwave_2021}.
An isolator is used at the input of the device, since its cryogenic return loss is rated for  \SI{-28}{dB} at our readout frequency, compared to our attenuators which are rated for \SI{-19}{dB} return loss.

We calibrate to a through line that bypasses the sample and package (see Fig.~\ref{fig:fridge-diagram}).
Based on the measured $|S_{21}|$ away from the resonant point in Fig.~\ref{fig:app-plot-S21mag}, we estimate a package loss of \SI{0.28}{dB}.
The measured transmission characteristics are listed in Table~\ref{tab:S21}.

\section{Effective Linewidth in the Presence of Mismatches}

\begin{figure}
	\includegraphics{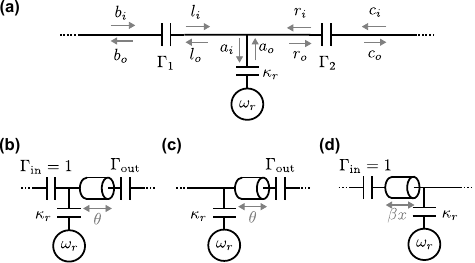}
	\caption{(a)~General input-output network of a resonator mode connected to feedline with reflection coefficients of $\Gl$ on the left and $\Gr$ on the right.
		(b)~and~(c)~Networks to analyze the variation in $\kr$ due to off-chip mismatches $\Gout$ with and without intentional mismatch ($\Gin=\{1,0\}$).
		(d)~Network to analyze the spread of $\kr$ due to on-chip non-uniformities in the presence of intentional mismatch $\Gin=1$ with electrical length $\beta x$.
	}
	\label{fig:app-kappa}
\end{figure}

Here, we derive the general expression for the effective linewidth $\kat$ of a single-mode resonator in the presence of mismatches at both ends of the feedline, using the input-output relations \cite{gardiner_input_1985}, also known as coupled mode theory \cite{haus_waves_1984}.
Since the effective linewidth $\kat$ is a function of the standing waves in the feedline between the mismatches, it is dependent on the reflection coefficients $\Gl$ and $\Gr$, which we capture by representative capacitances in this analysis.
This derivation expands on that in \cite{heinsoo_rapid_2018}, which calculated the linewidth of a resonant mode in the presence of capacitance on one end of the feedline.

We depict the input-output network in Fig.~\ref{fig:app-kappa}a. The evolution of the resonator mode $a$ is given by 
\begin{equation}
	\frac{d\amode}{dt}=j\wa\amode-\frac{\ka}{2}\amode+\sqrt{\ka}\ai.
	\label{eq:eom}
\end{equation}
The input-output relations of the resonator are governed by
\begin{equation}
	\aoo=\ai-\sqrt{\ka}\amode\,,
\end{equation}
with input ($i$) and output ($o$) directions shown in Fig.~\ref{fig:app-kappa}a.

The T-junction connecting the resonator to the feedline's right ($r$) and left ($l$) sides is symmetric, reciprocal, and lossless. 
The scattering relations that satisfy this are \cite{pozar_microwave_2012,heinsoo_rapid_2018}
\begin{equation}
	\lo=-\frac{1}{3}\li+\frac{2}{3}\ri+\frac{2}{3}\aoo\,,
\end{equation}
\begin{equation}
	\ro=\frac{2}{3}\li-\frac{1}{3}\ri+\frac{2}{3}\aoo\,,
\end{equation}
and
\begin{equation}
	\ai=\frac{2}{3}\li+\frac{2}{3}\ri-\frac{1}{3}\aoo.
\end{equation}
The input-side capacitance has the scattering relations \cite{pozar_microwave_2012}
\begin{equation}
	\co=\left(1-\Gl\right)\lo+\Gl\ci
\end{equation}
and
\begin{equation}
	\li=\left(1-\Gl\right)\ci+\Gl\lo.
\end{equation}
Similarly, the output-side capacitance has the scattering relations
\begin{equation}
	\doo=\left(1-\Gr\right)\ro+\Gr\di
\end{equation}
and
\begin{equation}
	\ri=\left(1-\Gr\right)c_i+\Gr\ro.
\end{equation}
Eliminating the $\li$ and $\lo$ modes, we obtain
\begin{equation}
	\ai=\frac{1}{2}\left(1-\Gl\right)\ci+\frac{1}{2}\left(1+\Gl\right)\ri+\frac{1}{2}\frac{\sqrt{\ka}}{2}\left(1-\Gl\right)\amode.
\end{equation}
Next, eliminating the $\ri$ and $\ro$ modes, we find
\begin{equation}
	\begin{aligned}
		\ai=&\frac{1}{2}\frac{\left(1+\Gl\right)\left(1-\Gr\right)}{1-\Gl\Gr}\di
		+\frac{1}{2}\frac{\left(1-\Gl\right)\left(1+\Gr\right)}{1-\Gl\Gr}\ci\\
		&+\frac{1}{2}\frac{\sqrt{\ka}}{2}\frac{1-\Gl-\Gr-3\Gl\Gr}{1-\Gl\Gr}\amode.
	\end{aligned}
\end{equation}
We can then substitute into the original equation of motion in \eqref{eq:eom} to obtain
\begin{equation}
	\begin{aligned}
		\frac{d\amode}{dt}&=j\wat\amode-\frac{\kat}{2}\amode\\
		&+\frac{\sqrt{\ka}}{2}\frac{\left(1+\Gl\right)\left(1-\Gr\right)}{1-\Gl\Gr}\di\\
		&+\frac{\sqrt{\ka}}{2}\frac{\left(1-\Gl\right)\left(1+\Gr\right)}{1-\Gl\Gr}\ci\,,
	\end{aligned}
\end{equation}
where the effective frequency is
\begin{equation}
	\wat=\wa+\frac{\ka}{4}\Imag{\frac{1-\Gl-\Gr-3\Gl\Gr}{1-\Gl\Gr}}\,,
\end{equation}
and the effective linewidth is
\begin{equation}
	\kat=\frac{\ka}{2}\Real{\frac{\left(1+\Gl\right)\left(1+\Gr\right)}{1-\Gl\Gr}}.
	\label{eq:effkappa}
\end{equation}
In the following sections, we use this general form to derive the expressions given in Sec.~\ref{sec:motivation}.

\subsection{\texorpdfstring{Spread Due to Off-Chip Mismatch:\\ With Intentional Mismatch}{Spread Due to Off-Chip Mismatch: With Intentional Mismatch}}
\label{sec:app-int-mismatch}
We consider the special case where $\Gl=\Gin=1$, representing intentional mismatch of total reflection with negligible dispersion.
The effective linewidth in \eqref{eq:effkappa} simplifies to
\begin{equation}
	\kat\left(\Gin=1\right)=\ka\left(\frac{1-\abs{\Gr}^2}{1-2\Real{\Gr}+\abs{\Gr}^2}\right).
\end{equation}
We model $\Gr$ as an off-chip mismatch positioned electrical length $\theta$ away with magnitude $\Gn$ (see Fig.~\ref{fig:app-kappa}b). If the transmission line is lossless, then $\Gr=\Gn e^{-2j\theta}$ \cite{pozar_microwave_2012}.
We now have
\begin{equation}
	\kat\left(\Gin=1\right)=\ka\frac{1-\Gn^2}{1-2\Gn\cos2\theta+\Gn^2}\,.
\end{equation}
The ratio of the maximum ($\theta=0$) and minimum ($\theta=\pi/2$) linewidths simplifies to
\begin{equation}
	\frac{\kmax}{\kmin}\left(\Gin=1\right)=\left(\frac{1+\Gn}{1-\Gn}\right)^2.
\end{equation}

\subsection{\texorpdfstring{Spread Due to Off-Chip Mismatch:\\ No Intentional Mismatch}{Spread Due to Off-Chip Mismatch: No Intentional Mismatch}}
\label{sec:app-noint-mismatch}
We now consider the case where $\Gl=\Gin=0$, which is characteristic of a feedline with a well-matched input side (see Fig.~\ref{fig:app-kappa}c).
The effective linewidth in \eqref{eq:effkappa} simplifies to
\begin{equation}
	\kat\left(\Gin=0\right)=\ka\left(\frac{1+\Real{\Gr}}{2}\right).
	\label{eq:tmp1}
\end{equation}
Substituting in $\Gr=\Gn e^{-2j\theta}$, \eqref{eq:tmp1} simplifies to
\begin{equation}
	\kat\left(\Gin=0\right)=\ka\left(\frac{1+\Gn\cos2\theta}{2}\right).
\end{equation}
The ratio of the maximum ($\theta=0$) and minimum ($\theta=\pi/2$) linewidths is given by
\begin{equation}
	\frac{\kmax}{\kmin}\left(\Gin=0\right)=\frac{1+\Gn}{1-\Gn}.
\end{equation}

\subsection{\texorpdfstring{Sensitivity to On-Chip Non-Uniformities:\\ With Intentional Mismatch}{Sensitivity to On-Chip Non-Uniformities: With Intentional Mismatch}}
\label{sec:app-nonuniform}
For a feedline with intentional mismatch, we derive the effect of non-uniformities on $\krt$. 
For simplicity, we omit the effect of off-chip mismatches on the output side and set $\Gr=0$.
We now have
\begin{equation}
	\kat=\ka\left(\frac{1+\Real{\Gl}}{2}\right).
\end{equation}
We assume the resonator is spaced some distance away from the intentional mismatch (see Fig.~\ref{fig:app-kappa}d).
The reflection coefficient of a lossless transmission line of length $x$ terminated in a mismatch with magnitude $\abs{\Gin}$ is given by \cite{pozar_microwave_2012}
\begin{equation}
	\Gl=\abs{\Gin}e^{-2j\beta x}\,,
\end{equation}
where ${\beta}$ is the phase constant.
In the case of intentional mismatch, $\Gin=1$.
We now have
\begin{equation}
	\krt(x)=\krn\left[\frac{1}{2}\cos\left(2{\beta} x\right)+\frac{1}{2}\right]\,,	
	\label{eq:kappastep2}
\end{equation}
where we have denoted $\krn=\kr$ to represent the maximum possible linewidth.
Assuming the readout tone is at the resonator frequency $\wrr$ with wavelength $\lr$, the phase constant $\beta$ along the feedline is given by
\begin{equation}
	\beta=\frac{\wrr}{\vpfl}=\frac{2\pi\vpr}{\lr\vpfl}=\frac{2\pi}{\lr}\sqrt{\frac{\efl}{\eres}}\,,
\end{equation}
where \{$\vpfl$, $\vpr$\} and \{$\efl$, $\eres$\} are the phase velocity and effective permittivity of the \{feedline, resonator\}.
For simplicity, we have not included the effects of $\eeff$ on $\krn$.
Substituting into  \eqref{eq:kappastep2}, we have
\begin{equation}
	\tilde{\kappa}_r\left(x\right) = \krn\left[\frac{1}{2}\cos\left(2\pi\frac{x}{\lr/2}\sqrt{\frac{\efl}{\eres}}\right) + \frac{1}{2}\right].
	\label{eq:kappafull}
\end{equation}

\section{Linewidth Variation Due to On-Chip Nonuniformities}
\begin{figure}
	\includegraphics{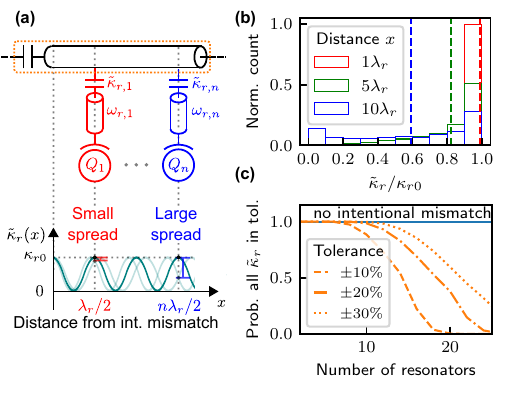}
	\caption{Sensitivity of resonator linewidth $\kr$ to on-chip non-uniformities resulting in $\efl\neq\eres$ error. Here, we do not include the additional variation that would result from off-chip mismatches. (a)~Illustration of spread in fabricated $\tilde{\kappa}_r$ as a function of $x$ in the presence of $\efl\neq\eres$. 
		(b)~Histograms of $\krt$ for increasing $x$, where the vertical dashed line denotes the target design value at the mean $\mu_\kappa/\krn$. We assume a normal distribution of $\sqrt{\efl/\eres}$ errors with a mean of 1 (no error) and a relative standard deviation of 1.5\% (e.g., a shift of \SI{120}{MHz} for an \SI{8}{GHz} resonator). 
		(c)~The probability that all fabricated resonators' $\krt$ are within tolerance of $\mu_\kappa$ as a function of the number of resonators. We assume two resonators are coupled every $\lr/2$, similar to the layout of \cite{heinsoo_rapid_2018}.}
	\label{fig:kappa-sensitivity}
\end{figure}

\label{sec:app-nonuniform-full}
Here, we perform a detailed analysis of the linewidth variation resulting from on-chip nonuniformities, in the presence of intentional mismatch.
We consider the circuit in Fig. ~\ref{fig:kappa-sensitivity}a, where multiplexed resonators are positioned along a feedline interrupted by intentional mismatch (e.g., an input capacitance). 
For a resonator at a distance $x$ from the intentional mismatch, the fabricated linewidth $\tilde{\kappa}_r$ is approximated by (see Appendix~\ref{sec:app-nonuniform}) 
\begin{equation}
	\tilde{\kappa}_r\left(x\right) = \krn\left[\frac{1}{2}\cos\left(2\pi\frac{x}{\lr/2}\sqrt{\frac{\efl}{\eres}}\right) + \frac{1}{2}\right]\,,
	\label{eq:kappa}
\end{equation}
where $\krn$ is the largest possible linewidth, $\lr$ is the resonator wavelength, $\efl$ is the effective permittivity of the feedline between the input capacitance and the resonator, and $\eres$ is the effective permittivity of the resonator.
To maximize coupling to the waveguide, we position resonators at multiples of half-wavelength, similar to $x=n\lambda_{r}/2$ \cite{heinsoo_rapid_2018}. 
As visualized in Fig.~\ref{fig:kappa-sensitivity}a, due to the standing wave formed by the intentional mismatch, $\krt$ becomes more sensitive to non-uniformities in permittivity ($\efl\neq\eres$) at farther distances $x$.
To quantify this effect,  we assume that the $\sqrt{\efl/\eres}$ error for each resonator obeys a normal distribution with a mean of 1 (no error) and a relative standard deviation of 1.5\% (e.g., a shift of \SI{120}{MHz} for an \SI{8}{GHz} resonator).
In Fig.~\ref{fig:kappa-sensitivity}b, we plot histograms of $\krt$ for increasing distances $x$.
In Fig.~\ref{fig:kappa-sensitivity}c, we plot the probability that all fabricated $\krt$ are within a given tolerance as a function of the number of resonators, assuming that two resonators are coupled at each half-wavelength $\lr/2$, similar to \cite{heinsoo_rapid_2018}.
Even assuming a generous tolerance range of $\pm30\%$ for $\krt$, reliable fabrication of more than 15 multiplexed resonators becomes impractical.
In reality, the situation is even worse due to added variation from off-chip mismatches.
Because flip-chip processors are an attractive platform for the further scaling of quantum computers, it will be critical to make the resonator linewidth insensitive to on-chip non-uniformities moving forward.
Removing the intentional mismatch would achieve this goal (see Fig.~\ref{fig:kappa-sensitivity}c).
Moreover, this would make quantum processor design far more modular, as flexible placement of resonator-qubit blocks is possible.

\section{Broadband Amplifier Impedance Matching}
\label{sec:app-amplifier-matching}
\begin{figure}
	\includegraphics{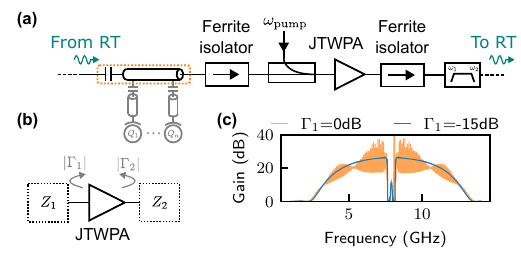}
	\caption{Importance of impedance matching for JTWPA performance. 
		(a)~Conventional quantum measurement chain with intentional mismatch.
		(b)~Circuit and (c)~gain curves to evaluate the effect of  mismatch on JTWPA gain performance.  We assume $\abs{\Gamma_2}=\SI{-15}{dB}$.
	}
	\label{fig:overhead}
\end{figure}
Here, we show that intentional mismatch adds to the overhead of multiplexed qubit readout by requiring nonreciprocal components to impedance match to downstream broadband amplifiers \cite{peng_x-parameter_2022, peng_floquet-mode_2022,heinsoo_rapid_2018}.
Contemporary quantum measurement chains use off-chip ferrite-based circulators and isolators biased by permanent magnets to realize wideband, unidirectional signal flow. 
While prevalent and essential for current experiments, these components act as a major inconvenience, since they dominate the available space at the base temperature stage and their high magnetic fields make integration with superconducting qubits infeasible.
While significant research has investigated the integration of on-chip circulators \cite{kamal_noiseless_2011,chapman_widely_2017,chapman_design_2019, ranzani_circulators_2019,kerckhoff_-chip_2015,sliwa_reconfigurable_2015}, no experimental devices with broad instantaneous bandwidth and sufficient isolation have yet been demonstrated.

We illustrate why broadband amplifier impedance matching is needed by considering a typical quantum measurement chain, shown in Fig.~\ref{fig:overhead}a.
Here, we use a Josephson traveling-wave parametric amplifier (JTWPA) for its broad bandwidth and high gain, which is suitable for multiplexed qubit readout \cite{macklin_near-quantum-limited_2015,obrien_resonant_2014}. 
We note that JTWPAs have the distinct advantage of transmission-based amplification, unlike Josephson parametric amplifiers (JPAs), which operate in reflection and inherently require circulators \cite{ranzani_circulators_2019}.

If not compensated, we show how intentional mismatch can create parametric oscillations in the JTWPA.
The environment surrounding the amplifier can be represented by the circuit in Fig.~\ref{fig:overhead}b.
We simulate the gain performance for different impedance environments using \texttt{JosephsonCircuits.jl} \cite{obrien_josephsoncircuitsjl_2023}. 
Our simulation uses a Floquet-mode JTWPA as the best-case scenario since it is inherently robust to out-of-band mismatch \cite{peng_floquet-mode_2022}.
For the output side, we assume $\abs{\Gamma_2}\approx\SI{-15}{dB}$, representing the reflection from the isolator at the JTWPA's output.
For the input side, we assume $\abs{\Gamma_1}\approx\{\SI{0}{dB},\SI{-15}{dB}\}$, representative of either the intentional mismatch or isolator at the input, respectively.
From Fig.~\ref{fig:overhead}c, we see that proper matching on the input side (i.e., low $\abs{\Gamma_1}$) strongly suppresses gain ripples, thus mitigating parametric oscillations and instability. We reiterate that this is a best-case analysis since the Floquet-mode JTWPA is robust to out-of-band mismatch.

The leading strategy of contemporary readout is to add an isolator before the JTWPA to compensate for the intentional mismatch \cite{peng_x-parameter_2022, peng_floquet-mode_2022,heinsoo_rapid_2018}.
We propose that, alternatively, one could instead make the quantum processor better impedance-matched by design by removing this mismatch.
This would reduce the need for isolation before the JTWPA; however, we note that isolation from the pump tone and amplified vacuum fluctuations would still be needed.
Nevertheless, removing the intentional mismatch would be a major step toward the monolithic integration of qubits and quantum-limited amplifiers.
Such integration would both drastically reduce the footprint of the external microwave infrastructure and mitigate preamplification losses.

\section{Input-Output Theory of All-Pass Readout}
\label{sec:app-cmt}
This appendix derives the $S_{21}$ expression in \eqref{eq:drr-S21} for a resonator with an even and odd mode, again using the coupled mode equations \cite{haus_waves_1984}. We also show the equivalence of the generalized system with the two-resonator system.

\begin{figure}
	\includegraphics{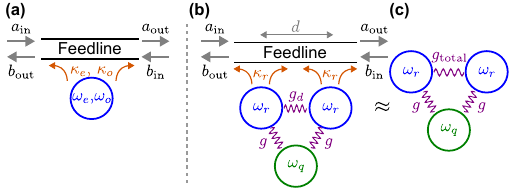}
	\caption{(a)~Generalized system with even and odd modes. (b)~Two resonators coupled to a feedline and a  qubit mode. (c)~Simplification of (b).}
	\label{fig:app-cmt}
\end{figure}

First, we consider a generalized resonator with an even and odd mode with amplitudes $\ahe$ and $\aho$, respectively, as in Fig.~\ref{fig:app-cmt}a. For such modes, we expect that forward and backward propagating waves would couple in phase to the symmetric mode and out of phase to the antisymmetric mode. We can determine the mode evolution as \cite{manolatou_coupling_1999}
\begin{equation}
	\frac{d\ahe}{d t}=\left(j \wre-\frac{\ke}{2}\right) \ahe+\sqrt{\frac{\ke}{2}}\left(\ain+\bin\right)
	\label{eq:2m1rdas}
\end{equation}
and
\begin{equation}
	\frac{d \aho}{d t}=\left(j \wro-\frac{\ko}{2}\right) \aho+\sqrt{\frac{\ko}{2}}\left(\ain-\bin\right).
	\label{eq:2m1rdaa}
\end{equation}
The input-output equations are given by
\begin{equation}
	\aout=\ain-\sqrt{\frac{\ke}{2}}\ahe-\sqrt{\frac{\ko}{2}}\aho
\end{equation}
and
\begin{equation}
	\bout=\bin-\sqrt{\frac{\ke}{2}} \ahe+\sqrt{\frac{\ko}{2}}\aho.
\end{equation}
Letting $\bin = 0$, we can solve for the transmission $S_{21}$ as
\begin{equation}
	\begin{aligned}
		S_{21} = \frac{\aout}{\ain}=&1-\frac{\frac{\ke}{2}}{j\left(\omega-\omega_{e}\right)+\frac{\ke}{2}}-\frac{\frac{\ko}{2}}{j\left(\omega-\omega_{o}\right)+\frac{\ko}{2}}.
	\end{aligned}
	\label{eq:2m1rS21}
\end{equation}
If the modes are degenerate, then they have equal frequencies and equal linewidths, i.e.,
\begin{equation}
	\omega_e=\omega_o=\omega_{\text{deg}}
\end{equation}
and
\begin{equation}
	\kappa_{e} = \kappa_{o} = \kappa_{\text{deg}}.
\end{equation}
Eq. \eqref{eq:2m1rS21} then becomes
\begin{equation}
	S_{21,\mathrm{deg}} = 1-\frac{\kappa_\text{deg}}{j\left(\omega-\omega_{\text{deg}}\right)+\frac{\kappa_\text{deg}}{2}}.
\end{equation}
This is the desired all-pass behavior, since we have $|S_{21,\mathrm{deg}}|=1$ at $\omega=\omega_\mathrm{deg}$.

Now, we consider our system, as shown in Fig.~\ref{fig:app-cmt}b. We approximate the weakly anharmonic transmon qubit as a linear oscillator. 
This analysis derives the waveguide-mediated coupling and $\kappa_e=\kappa_o$ matching condition but will omit the nonlinear correction for $\wre$ and $\wro$. 
See Appendix~\ref{sec:app-hamiltonian} for the calculation of $\wre$ and $\wro$ with the nonlinear correction. 
From coupled-mode theory \cite{haus_waves_1984}, the evolution of the modes is given by
\begin{equation}
	\begin{aligned}
		\frac{d \aha}{d t}&=\left(j \omega_{r}-\frac{\kr}{2}\right) \aha+j g_d \ahb + jg\bh
		+\sqrt{\frac{\kr}{2}} \ain\\
		&+\sqrt{\frac{\kr}{2}} e^{-j\phi}  \cdot\left(\bin-\sqrt{\frac{\kr}{2}}\ahb\right)\,,
	\end{aligned}
	\label{eq:2m2raL}
\end{equation}
\begin{equation}
	\begin{aligned}
		\frac{d \ahb}{d t}&=\left(j \omega_{r}-\frac{\kr}{2}\right) \ahb+j g_d \aha +jg\bh
		+\sqrt{\frac{\kr}{2}} \bin\\
		&+\sqrt{\frac{\kr}{2}} e^{-j\phi}  \cdot\left(\ain-\sqrt{\frac{\kr}{2}}  \aha\right)\,,
	\end{aligned}
	\label{eq:2m2raR}
\end{equation}
and
\begin{equation}
	\frac{d\bh}{d t}=j\wq\bh + jg\aha + jg\ahb.
	\label{eq:2m2b}
\end{equation}
where $\phi = \beta d$, $\beta$ is the propagation constant, and $d$ is the separation between the resonators.

Assuming $e^{j\omega t}$ dependence, we can solve for a steady-state expression for the qubit mode $\bh$, where \eqref{eq:2m2b} becomes
\begin{equation}
	\bh(\omega) = \frac{g}{\omega-\wq}\left(\aha+\ahb\right).
	\label{eq:2m2bsol}
\end{equation}
Noting that the qubit mode mediates a coupling at $\omega=\wrr$, we substitute \eqref{eq:2m2bsol} into \eqref{eq:2m2raL} and \eqref{eq:2m2raR}, and obtain
\begin{equation}
	\begin{aligned}
		\frac{d}{d t} \ahab &=\left[j \left(\wrr-\frac{g^2}{\wq-\wrr}\right)-\frac{\kr}{2}\right] \ahab\\&+j \left(g_d-\frac{g^2}{\wq-\wrr}\right) \ahba\\&
		+\sqrt{\frac{\kr}{2}} \ainbin\\&
		+\sqrt{\frac{\kr}{2}} e^{-j\phi}  \left(\binain
		-\sqrt{\frac{\kr}{2}}\ahba\right).
	\end{aligned}
	\label{eq:2m2rv2}
\end{equation}
Defining the even and odd modes as	
\begin{equation}
	\aheo=\frac{1}{\sqrt{2}}\left\{\begin{array}{c}
		\aha+\ahb \\
		\aha-\ahb
	\end{array}\right\}\,,
	\label{eq:transform}
\end{equation}
we can derive the evolution of the even and odd modes (without nonlinear correction) analogous to \eqref{eq:2m1rdas} and \eqref{eq:2m1rdaa} as
\begin{equation}
	\begin{aligned}
		\frac{d}{d t}\aheo&= 
		\left\{\begin{array}{l}
			j\wre-\frac{\ke}{2} \\
			j\wro-\frac{\ko}{2}
		\end{array}\right\} \aheo \\
		&+\sqrt{{\kr}} e^{-j\phi / 2}\left\{\begin{array}{c}
			\cos \frac{\phi}{2} \\
			j \sin \frac{\phi}{2}
		\end{array}\right\}
		\left\{\begin{array}{c}
			\ain+\bin\\
			\ain-\bin
		\end{array}\right\}\,, \\
	\end{aligned}
	\label{eq:2m2rasaa}
\end{equation}
where
\begin{equation}
	\left\{\begin{array}{l}
		\omega_e \\
		\omega_o
	\end{array}\right\}=\left\{\begin{array}{l}
		\wrr+\gfix-\frac{2g^2}{\wq-\wrr} \\
		\wrr-\gfix
	\end{array}\right\}\,,\\
	\label{eq:2m2rwewo}
\end{equation}
\begin{equation}
	\left\{\begin{array}{l}
		\ke \\
		\ko
	\end{array}\right\}=\kr\left\{\begin{array}{c}
		1+\cos\phi \\
		1-\cos\phi
	\end{array}\right\}\,,\\
\end{equation}
and
\begin{equation}
	\gfix=g_d+g_w=g_d+\frac{\kr}{2}\sin\phi.
\end{equation}
The waveguide-mediated coupling is thus given by $g_w=\frac{\kr}{2}\sin\phi$. To match the linewidths of the even and odd modes $\ke=\ko$, we require the electrical length to satisfy $\phi=\pi(m+1)/2$.

\section{Hamiltonian of All-Pass Readout}
\label{sec:app-hamiltonian}

This appendix derives the Hamiltonian for the circuit in Fig.~\ref{fig:app-cmt}c. The Hamiltonian for two resonators coupled symmetrically to a qubit in the transmon regime is approximated in the Fock basis by
\begin{equation}
	\begin{aligned}
		{H}  &=  \wrr\ahda\aha + \wrr\ahdb\ahb + \wq\bhd\bh - \frac{E_C}{2}\bhd\bhd\bh\bh \\
		&+ \gfix\left(\ahda+\aha\right)\left(\ahdb+\ahb\right) \\
		&+ g\left(\ahda+\aha\right)\left(\bhd+\bh\right)
		+ g\left(\ahdb+\ahb\right)\left(\bhd+\bh\right).
	\end{aligned}
	\label{eq:symHfull}
\end{equation}
We now transform the basis to normal, diagonalized modes of the system. Since the resonators are symmetrically coupled to the qubit, we can transform to an even and odd basis by the transformation in \eqref{eq:transform}, which is equivalently,
\begin{equation}
	\ahab=\frac{1}{\sqrt{2}}\left\{\begin{array}{c}
		\ahe+\aho \\
		\ahe-\aho
	\end{array}\right\}.
	\label{eq:transform2}
\end{equation}
Applying the rotating wave approximation to \eqref{eq:symHfull} and substituting the transformation from \eqref{eq:transform2}, we obtain
\begin{equation}
	\begin{aligned}
		{H}
		&= \wrr\ahde\ahe + \wrr\ahdo\aho + \wq\bhd\bh -\frac{E_C}{2}\bhd\bhd\bh\bh \\
		&+\gfix\left(\ahde\ahe-\ahdo\aho\right) 
		+ \sqrt{2}g\left(\ahde\bh+\ahe\bhd\right).
	\end{aligned}
\end{equation}
Only the last term needs to be diagonalized. Assuming this interaction term is small relative to the detuning, we can perform a standard Schrieffer-Wolff transformation \cite{blais_circuit_2021}, truncate to the first two levels, and derive the Hamiltonian in the dispersive limit as
\begin{equation}
	\begin{aligned}
		{\tilde{H}}_{\mathrm{disp}}
		& = \frac{\wqt}{2}\sz + \wre \ahde\ahe + \wro\ahdo\aho\,,
	\end{aligned}
\end{equation}
where the dressed qubit frequency $\wqt$, even mode frequency $\wre$, odd mode frequency $\wro$, and effective total dispersive
shift $2\chi_{01}$ are given by
\begin{equation}
	\wqt=\wq+\frac{2g^2}{\Delta}\,,
\end{equation}
\begin{equation}
	\wre=\wrr+\gfix-\frac{2g^2}{\Delta-E_C}+2\chi_{01}\sz\,,
	\label{eq:wre}
\end{equation}
\begin{equation}
	\wro=\wrr-\gfix\,,
	\label{eq:wro}
\end{equation}
and
\begin{equation}
	\chi_{01}=-\frac{g^2E_C}{\Delta\left(\Delta-E_C\right)}\,,
\end{equation}
where the qubit-resonator detuning is $\Delta=\wq-\wrr$.

\section{\texorpdfstring{Scaling of $\mathbf{2\chi_{01}/\kr}$}{Scaling of 2Chi/Kappa}}
\label{sec:app-chikappa}
\begin{figure}[t]
			\includegraphics[width=\columnwidth]{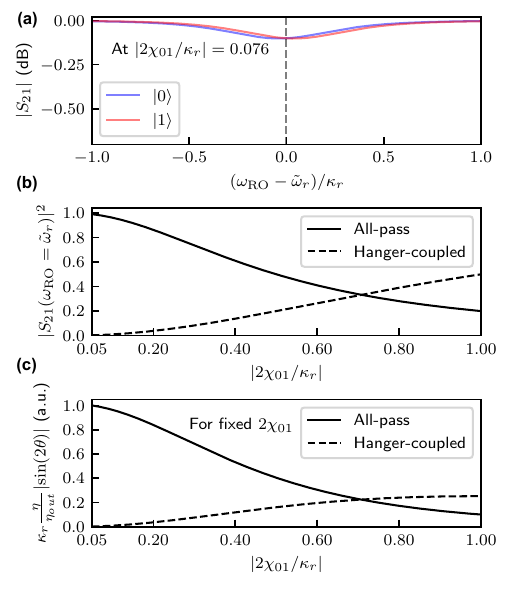}
	\caption{
		(a)~Analytic $|S_{21}|$ versus $\wRO$ for $|2\chi_{01}/\kr|=0.076$ .
		(b)~Analytic $|S_{21}|$ versus $|2\chi_{01}/\kr|$ for the all-pass versus hanger-coupled resonator with no intentional mismatch. The readout tone is positioned at $\wRO=\wrt$. 
			(c)~Comparison of $\kr\frac{\eta}{\etaout}|\sin(2\theta)|$ for fixed $2\chi_{01}$.
}
	\label{fig:app-chi-kappa-combined}
\end{figure}
Due to the qubit's nonlinearity, the degeneracy of our system will be disrupted depending on the qubit state. Here, we discuss how the transmission of this form of all-pass readout is expected to scale with $2\chi_{01}/\kr$. We assume the case in which $\gfix=g^2/(\Delta-E_C)$ so that the even and odd modes are roughly equal $\wrt=\wro\approx\wre$. We denote the dressed resonator mode as $\wrt$. In this case, from \eqref{eq:wre} and \eqref{eq:wro}, we obtain
\begin{equation}
	\wre=\wrt+2\chi_{01}\sz
\end{equation}
and
\begin{equation}
	\wro=\wrt.
\end{equation}
We also assume a lossless feedline and that the even and odd modes have equal linewidths $\kr=\ke=\ko$. We substitute into \eqref{eq:2m1rS21} and have
\begin{equation}
	\begin{aligned}
		\left\{\begin{array}{l}
			S_{21,\ground} \\
			S_{21,\excited}
		\end{array}\right\} = 1&-\frac{\frac{\kr}{2}}{j\left(\wRO-\wrt\pm2\chi_{01}\right)+\frac{\kr}{2}}\\
		&-\frac{\frac{\kr}{2}}{j\left(\wRO-\wrt\right)+\frac{\kr}{2}},
	\end{aligned}
\label{eq:S21ge}
\end{equation}
where $\wRO$ is the readout frequency.
Using \eqref{eq:S21ge}, we can determine the effective total dispersive shift to be $2\chieff=2\chi_{01}$, as it satisfies $S_{21,\ground}\left(\wRO\right)=S_{21,\excited}\left(\wRO+2\chieff\right)$.

The maximum difference in phase between the two states occurs for a readout tone positioned at $\wRO=\wrt$. The transmission at this frequency is given by
\begin{equation}
	\left\{\begin{array}{l}
		S_{21,\ground} \\
		S_{21,\excited}
	\end{array}\right\}\left(\wRO=\wrt\right) = -\frac{\frac{\kr}{2}}{\frac{\kr}{2}\pm j2\chi_{01}}.
\label{eq:S21chikappa}
\end{equation}
The magnitude $|S_{21}|$ is the same for both the ground and excited states and is plotted versus $|2\chi_{01}/\kappa_r|$ in Fig.~\ref{fig:app-chi-kappa-combined}b.
The proportion of measurement photons decaying towards the output is characterized by $|S_{21}|^2$.
Targeting large $|S_{21}|^2\sim 0.98$, we operate at $|2\chi_{01}/\kappa_r|=0.076$, denoted by the vertical dashed line.
For this parameter choice, we plot the analytic $|S_{21}|$ versus $\wRO$ in Fig.~\ref{fig:app-chi-kappa-combined}a.

We provide comparison to a hanger-coupled resonator without intentional mismatch.
	The transmission of a hanger-coupled resonator has the standard  form given by \cite{haus_waves_1984, blais_circuit_2021}
	\begin{equation}
		\left\{\begin{array}{l}
			S_{21,\ground} \\
			S_{21,\excited}
		\end{array}\right\} = 1-\frac{\frac{\kr}{2}}{j\left(\wRO-\wrt\pm\chi_{01}\right)+\frac{\kr}{2}}
		\label{eq:S21ge_std}
	\end{equation}
	If we perform readout by probing at $\wRO=\wrt$, we find that the transmission is given by
	\begin{equation}
		\left\{\begin{array}{l}
			S_{21,\ground} \\
			S_{21,\excited}
		\end{array}\right\}\left(\wRO=\wrt\right) = 1-\frac{\frac{\kr}{2}}{\frac{\kr}{2}\pm j\chi_{01}}.
		\label{eq:S21chikappa_std}
	\end{equation}
In Fig.~\ref{fig:app-chi-kappa-combined}b, we compare the $|S_{21}(\wRO=\wrt)|^2$ for our implementation of an all-pass resonator using \eqref{eq:S21chikappa} compared with a hanger-coupled resonator using \eqref{eq:S21chikappa_std}.

We perform a comparison of the signal-to-noise ratio (SNR) of these two approaches, using the standard expression given by \cite{blais_circuit_2021,bultink_general_2018} 
\begin{equation}
\text{SNR}^2 \propto\eta\kr\bar{n}|\sin(2\theta)|t
	=\etaout\bar{n}\left[\kr\frac{\eta}{\etaout}|\sin(2\theta)|\right]t
, 
\label{eq:snr}
\end{equation}
where $\sin(2\theta)=\chi_{01}\kr/(\chi_{01}^2+\kr^2/4)$ and $\etaout$ is the efficiency of the output chain.
We make two observations.
First, without intentional mismatch, the ratio $\frac{\eta}{\etaout}$ is equivalent to $|S_{21}(\wRO=\wrt)|^2$, as previously shown in Fig.~\ref{fig:app-chi-kappa-combined}b.
Second, in practice, the total dispersive shift for state-of-the-art transmon qubits with low anharmonicity is limited to $|2\chi_{01}/2\pi|\lesssim\SI{20}{MHz}$ since the perturbative dispersive interaction is constrained to a small fraction of the qubit anharmonicity  \cite{walter_rapid_2017,koch_charge-insensitive_2007}.
As a result, the leading approach to achieve larger SNR and faster readout has been to use larger $\kr$ and correspondingly lower $2\chi_{01}/\kr$ (i.e., increasing $\kr$ instead of $|\sin(2\theta)|$, which is at most 1) \cite{walter_rapid_2017,sunada_fast_2022}.
A recent work achieved readout fidelity of 99.1\% in 40 ns using a ratio of $|2\chi_{01}/\kr|\approx0.15$ \cite{sunada_fast_2022}.
Taking into account these practical considerations, we can compare the SNR of the all-pass and hanger-coupled resonators by plotting $\kr\frac{\eta}{\etaout}|\sin(2\theta)|$ as a function of $|2\chi_{01}/\kr|$ (see Fig.~\ref{fig:app-chi-kappa-combined}c), where $2\chi_{01}$ is fixed to represent a practical upper limit.
While the hanger geometry without intentional mismatch reaches its maximum value at $|2\chi_{01}/\kr|=1$, the all-pass resonator exceeds that maximum for $|2\chi_{01}/\kr|<0.67$ and more than doubles it for $|2\chi_{01}/\kr|<0.42$.

	Finally, we comment on the experimental results, where the resonator linewidth $\kr/2\pi=\SI{14.5}{MHz}$ is held constant, in contrast to Fig.~\ref{fig:app-chi-kappa-combined}c.
	Shelving increases the magnitude of the effective total dispersive shift from \SI{1.10}{MHz} to \SI{1.86}{MHz}.
	This increases $\frac{\eta}{\etaout}|\sin(2\theta)|$ from 0.14 to 0.23, resulting in an SNR improvement.

\section{Frequency Spacing}
\label{sec:app-multiplex}
\begin{figure}[t]
	\includegraphics{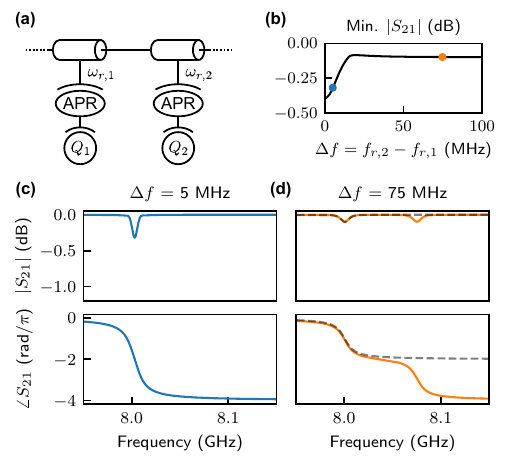}
	\caption{
		Analytic transmission of two cascaded all-pass readout units. 
		(a)~Two cascaded all-pass readout units.
		(b)~Minimum $|S_{21}|$ as a function of frequency spacing.
		Transmission for (c)~$\Delta f=\SI{5}{MHz}$ and (d)~$\Delta f=\SI{75}{MHz}$ are plotted for comparison.
		We assume $f_{r,1}=\SI{8}{GHz}$, $\kr/2\pi=\SI{15}{MHz}$, and $|2\chi_{01}/\kappa|=0.076$.
		In (d), the gray dashed curve corresponds to the transmission of a single all-pass readout unit.
	}
	\label{fig:app-spacing}
\end{figure}

Here, we analyze the effect of frequency spacing on the directionality of multiplexed all-pass readout resonators.
We consider two cascaded readout units as in Fig.~\ref{fig:app-spacing}a, assuming parameters of $f_{r,1}=\SI{8}{GHz}$, $\kr/2\pi=\SI{15}{MHz}$, and $|2\chi_{01}/\kappa|=0.076$.
For simplicity, we assume the qubit is in the ground state and that $\wrt=\omega_o$, as done in Appendix~\ref{sec:app-chikappa}.
In Fig.~\ref{fig:app-spacing}b, we plot the minimum transmission as a function of the frequency spacing $\Delta f=f_{r,2}-f_{r,1}$.
We see that for frequency spacing less than the resonator linewidth (e.g., Fig.~\ref{fig:app-spacing}c), directionality is degraded due to the interaction of resonators between the different readout units.
For frequency spacing greater than the resonator linewidth (e.g., Fig.~\ref{fig:app-spacing}d), the directionality is unaffected.

\section{Finite-Element Simulation}
\label{sec:app-fem}

\begin{figure}[t]
	\includegraphics{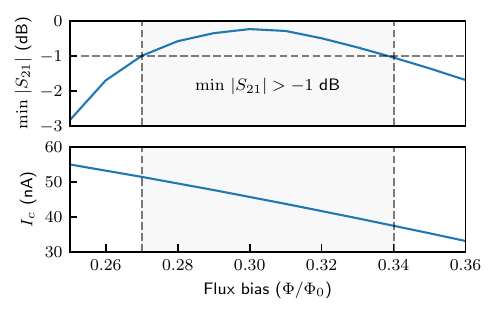}
	\caption{
			(a)~Minimum $|S_{21}|$ from finite-element simulation as a function of flux bias. 
			(b)~Corresponding Josephson inductance as a function of flux bias.
			The shaded region denotes the range of values where the minimum $|S_{21}|>\SI{-1}{dB}$.
	}
	\label{fig:plot-S21-min}
\end{figure}

We predict the performance of our device using  finite-element method (FEM) simulation in Ansys High Frequency Simulation Software (HFSS). We first fine-tune the permittivity to match the measured resonator frequency. We model our system as a 3-port device, where ports 1 and 2 are the readout input and output, respectively, and port 3 replaces the Josephson junction.

As shown in Fig.~\ref{fig:drr_S21}a, we obtain agreement between simulated and experimental $|S_{21}|$ as a function of the flux biasing the SQUID of the flux-tunable transmon. 
Approximating the transmon qubit as a linear oscillator, we terminate port 3 with an inductor and simulate $S_{21}$. 
We sweep the inductance according to the Josephson inductance of a symmetric SQUID given by $L_J(\Phi)=\frac{1}{E_J(\Phi)}\left(\frac{\hbar}{2e}\right)^2$, where $E_J(\Phi)=2E_{J}\left|\cos(\pi\frac{\Phi}{\Phi_0})\right|$ \cite{blais_circuit_2021} with fitted $E_{J}=\SI{19.3}{GHz}$. 
We plot the minimum transmission as a function of flux bias in Fig.~\ref{fig:plot-S21-min}a and show corresponding critical current  $I_c=\frac{2\pi}{\Phi_0}E_J(\Phi)$ in Fig.~\ref{fig:plot-S21-min}b.
We see that to achieve transmission above \SI{-1}{dB}, critical currents in the range of \SI{37.4}{nA} to \SI{51.4}{nA} are acceptable (i.e., for target $I_c=\SI{43.8}{nA}$, the tolerance range is $\pm17\%$).
Since fluctuations in $I_c$ of a few percent are typical for modern fabrication processes \cite{kreikebaum_improving_2020,pop_fabrication_2012}, this makes the proposed scheme compatible with fixed-frequency qubit architectures.

Next, we terminate both ports 1 and 2 with \SI{50}{\ohm} and perform eigenmode simulation to predict the linewidths of the two resonator modes of our system. Assuming the qubit is far detuned from the resonators, we predict linewidths of \SI{18.9}{MHz} and \SI{13.6}{MHz}, in agreement within 10\% with the experimental fits of \SI{17.1}{MHz} and \SI{11.9}{MHz}.

\section{Purcell Decay}
\begin{figure}[t!]
	\includegraphics{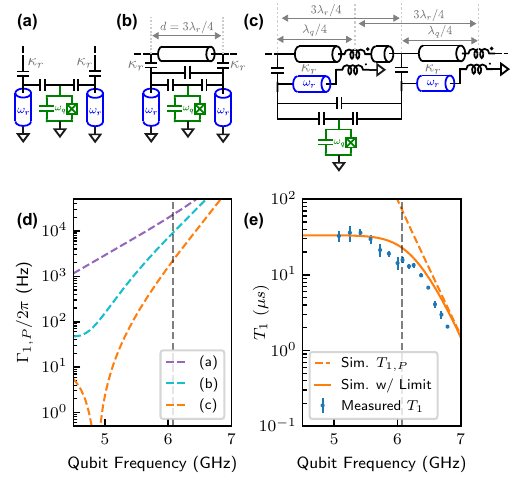}
	\caption{
		FEM simulation of Purcell decay.
		Corresponding circuit models of a qubit coupled to
		(a)~two resonators with separate loss channels,
		(b)~two resonators with shared feedline, where each resonator is coupled capacitively on one end,
		and (c)~two resonators with shared feedline, where each resonator is coupled to the feedline at two ends, corresponding to our tested device. 
		(d)~Simulated Purcell decay for (a)-(c).
		The dashed vertical line indicates the all-pass operating point in the main text.
		(e)~Measured and predicted qubit lifetime. We include a curve assuming a $T_1$ limit of \SI{33}{\micro s} due to intrinsic decay and the drive line.
	}
	\label{fig:plot-ipf}
\end{figure}

\label{sec:app-ipf}
Here, we simulate the Purcell decay using the same finite-element model as in Appendix~\ref{sec:app-fem}.
If we approximate the transmon qubit as a linear oscillator, we can calculate the qubit's Purcell decay as \cite{esteve_effect_1986,neeley_transformed_2008}
\begin{equation}
	\Gp=\frac{1}{\Tp}=\frac{\mathrm{Re}\left[Y_\mathrm{in}\left(\wq\right)\right]}{C_\Sigma}\,,
	\label{eq:Gp}
\end{equation}
where $Y_\mathrm{in}$ is the input admittance from the perspective of the Josephson junction and $C_\Sigma$ is the total capacitance of the transmon. 
We simulate the admittance parameters of this 3-port system, calculate the input admittance looking into port 3, and substitute into \eqref{eq:Gp}.

Using finite-element simulation, we simulate the Purcell decay of a qubit in three distinct configurations, shown in Fig.~\ref{fig:plot-ipf}a-c. 
The baseline model consists of a qubit coupled to two resonators with separate loss channels of decay rate $\kr$ (Fig.~\ref{fig:plot-ipf}a).
We then quantify the Purcell suppression achieved by two different configurations of all-pass readout.
The first couples each resonator to the shared feedline at one point with capacitive coupling (Fig.~\ref{fig:plot-ipf}b).
The second, corresponding to the measured device, couples each resonator to the shared feedline at two points with capacitive and inductive coupling (Fig.~\ref{fig:plot-ipf}c).

The simulated Purcell decay of the three configurations are plotted in Fig.~\ref{fig:plot-ipf}d.
To isolate the effect of Purcell decay into the feedline, these simulations do not include the dedicated drive line.
At the all-pass operating point of $\wq/2\pi=\SI{6080}{MHz}$ (denoted by the gray vertical line), the Purcell suppression obtained by using a shared feedline (Fig.~\ref{fig:plot-ipf}a$\to$b) is $\sim\!\!2.9$.
By coupling each resonator twice to the feedline instead of once (Fig.~\ref{fig:plot-ipf}b$\to$c), we can gain another factor of $\sim\!\!3.5$.
In total, this corresponds to a total suppression factor of $\sim\!\!10$ for the device in this work.

The suppression can be increased by aligning the notch frequency of the interferometric Purcell filter with the target $\wq$ used for all-pass readout.
From Fig.~\ref{fig:plot-ipf}d, the simulated Purcell suppression exceeds three orders of magnitude at the filter's notch frequency near \SI{4900}{MHz}, similar in magnitude to a separate demonstration of interference-based suppression in \cite{yen_interferometric_2024}.
The measured qubit lifetimes are shown in Fig.~\ref{fig:plot-ipf}e.
For comparison, we plot the Purcell-limited lifetime from simulation assuming a $T_1$ limit due to drive line and intrinsic losses.

\bibliography{main.bib}

\end{document}